\documentclass[12pt,twoside,prd,amsmath,amssymb,tightenlines,showkeys,
showpacs,eqsecnum]{revtex4}
\usepackage{epsfig}
\usepackage{mathptmx}
\usepackage{hyperref}
\usepackage{bm}
\usepackage{graphicx}
\newcommand{\hnl}{\htmladdnormallink}

\renewcommand{\cal}{\mathcal}

\newcommand {\ve}{\varepsilon}

\newcommand {\cG}{\cal G}
\newcommand {\cD}{\cal D}
\newcommand {\cL}{\cal L}

\newcommand {\bp}{\bar \psi}

\newcommand {\vf}{\varphi}

\def \myfigures #1#2#3#4#5#6#7#8
{\begin{figure}[ht]
    \begin{center}
        \includegraphics[width=#2 \textwidth]{#1.eps}
        \hfill
        \includegraphics[width=#6 \textwidth]{#5.eps}
        \parbox[t]{#4\textwidth}{\caption {#3}\label{#1}}
        \hfill
        \parbox[t]{#8\textwidth}{\caption {#7}\label{#5}}
    \end{center}
\end{figure} }

\def \myfigs #1#2#3#4#5#6#7#8
{\begin{figure}[ht]
    \begin{center}
        \includegraphics[width=#2 \textwidth, angle=-90]{#1.eps}
        \hfill
        \includegraphics[width=#6 \textwidth, angle=-90]{#5.eps}
        \vskip 5 mm
        \parbox[t]{#4\textwidth}{\caption {#3}\label{#1}}
        \hfill
        \parbox[t]{#8\textwidth}{\caption {#7}\label{#5}}
    \end{center}
\end{figure}}

\def \myfigss #1#2#3#4#5#6#7#8
{\begin{figure}[ht]
    \begin{center}
        \includegraphics[width=#2 \textwidth, angle=-90]{#1.eps}
        \hfill
        \includegraphics[width=#6 \textwidth]{#5.eps}
        \vskip 5 mm
        \parbox[t]{#4\textwidth}{\caption {#3}\label{#1}}
        \hfill
        \parbox[t]{#8\textwidth}{\caption {#7}\label{#5}}
    \end{center}
\end{figure}}

\def\myfigure #1#2#3#4
{\begin{figure}[ht]\begin{center}
\includegraphics[width=#2 \textwidth]{#1.eps}
\parbox[t]{#4\textwidth}{\caption{#3}\label{#1}}
\end{center}\end{figure}}

\begin{document}
\title{Nonlinear spinor field in Bianchi type-I cosmology: accelerated regimes}
\author{Bijan Saha}
\affiliation{Laboratory of Information Technologies\\ Joint
Institute for Nuclear Research, Dubna\\ 141980 Dubna, Moscow
region, Russia} \email{saha@thsun1.jinr.ru}
\homepage{http://thsun1.jinr.ru/~saha}
\date{\today}
\begin{abstract}
A self-consistent system of interacting nonlinear spinor and
scalar fields within the scope of a Bianchi type-I cosmological
model filled with perfect fluid is considered. Exact self-consistent 
solutions to the corresponding field equations are obtained. The role of spinor
field in the evolution of the Universe is studied. It is shown that the
spinor field gives rise to an accelerated mode of expansion of the Universe. 
At the early stage of evolution the spinor field nonlinearity generates the
acceleration while at the later stage it is done by the nonzero spinor mass.

\end{abstract}



\maketitle

\bigskip
\section{Introduction}

The accelerated mode of expansion of the present day Universe
encourages many researchers to introduce different kind of sources
that is able to explain this. Among them most popular is the dark
energy given by a $\Lambda$ term \cite{PRpadma,sahni,lambda},
quintessence \cite{caldwell,starobinsky,zlatev,pfdenr}, Chaplygin
gas \cite{kamen,pfden}. Recently cosmological models with spinor
field have been extensively studied by a number of authors in a
series of papers \cite{sgrg,sjmp,smpla,sprd,bited,green}. The
principal motive of the papers \cite{sgrg,sjmp,smpla,sprd,bited}
was to find out the regular solutions of the corresponding field
equations. In some special cases, namely with a cosmological
constant ($\Lambda$ term) that plays the role of an additional
gravitation field, we indeed find singularity-free solutions. It
was also found that the introduction of nonlinear spinor field
results in a rapid growth of the Universe. This allows us to
consider the spinor field as a possible candidate to explain the
accelerated mode of expansion. Note that similar attempt is made
in a recent paper by Kremer {\it et. al.} \cite{kremer1}. In this
paper we study the role of a spinor field in generating an accelerated
mode of expansion of the Universe. Since similar systems, though
from different aspects were thoroughly studied in \cite{sprd,bited},
to avoid lengthy calculations regarding spinor and scalar fields, we
mainly confine ourselves to the study of master equation
describing the evolution of BI Universe. We here give the
solutions to the spinor and scalar field equations, details of these
solutions can be found in \cite{sprd,bited}.

\section{Basic equations: a brief journey}

We consider a self consistent system of nonlinear spinor and
scalar fields within the scope of a Bianchi type-I gravitational
field filled with a perfect fluid. The spinor and the scalar field
is given by the Lagrangian
\begin{equation}
\cL = \frac{i}{2} \biggl[\bp \gamma^{\mu} \nabla_{\mu} \psi-
\nabla_{\mu} \bar \psi \gamma^{\mu} \psi \biggr] - m\bp \psi + F +
\frac{1}{2} (1 + \lambda_1 F_1)\,\vf_{,\alpha}\vf^{,\alpha},
\label{lag}
\end{equation}
where $\lambda_1$ is the coupling constant and $F$ and $F_1$ are
some arbitrary functions of invariants generated from the real
bilinear forms of a spinor field. Here we assume $F = F (I,J)$ and
$F_1 = F_1 (I,J)$ with $I = S^2$,\, $ S = \bp \psi$,\, $J = P^2$
and $P = i \bar \psi \gamma^5 \psi$.

The gravitational field is chosen in the form
\begin{equation}
ds^2 = dt^2 - a_1^2 dx_1^2 - a_2^2 dx_2^2 - a_3^2 dx_3^2,
\label{BI1}
\end{equation}
where $a_i$ are the functions of $t$ only and the speed of light
is taken to be unity. We also define
\begin{equation}
\tau = a_1 a_2 a_3. \label{taudef}
\end{equation}

We consider the spinor and scalar field to be space independent.
In that case for the spinor and the scalar fields and metric functions we find
the following expressions \cite{bited}.

For $F=F(I)$ we find $S=C_0/\tau$ with $C_0$ being an integration
constant. The components of the spinor field in this case read
\begin{equation}
\psi_{1,2}(t) = (C_{1,2}/\sqrt{\tau}) e^{-i\beta}, \quad
\psi_{3,4}(t) = (C_{3,4}/\sqrt{\tau}) e^{i\beta}, \label{psiI}
\end{equation}
with the integration constants obeying $C_0$ as $C_0 = C_{1}^{2} +
C_{2}^{2} - C_{3}^{2} - C_{4}^{2}.$ Here $\beta = \int(m -
{\cD})dt $ with ${\cD} = dF/dS + (\lambda_1 \dot{\vf}^2/2) dF_1/dS$.

For $F=F(J)$ in case of massless spinor field we find $P=D_0/\tau$. The
corresponding components of the spinor field in this case read:
with
\begin{eqnarray}
\psi_{1,2} &=& \bigl(D_{1,2} e^{i \sigma} + iD_{3,4}
e^{-i\sigma}\bigr)/\sqrt{\tau},\nonumber \\ \label{psiJ}\\ \psi_{3,4} &=&
\bigl(iD_{1,2} e^{i \sigma} + D_{3,4} e^{-i
\sigma}\bigr)/\sqrt{\tau},\nonumber
\end{eqnarray}
with $D_0=2\,(D_{1}^{2} + D_{2}^{2} - D_{3}^{2} -D_{4}^{2}).$ Here
$\sigma = \int {\cG} dt$ with ${\cG} = dF/dP + (\lambda_1 \dot{\vf}^2/2) dF_1/dP$.

For the scalar field we find
\begin{equation}
\vf = C \int \frac{dt}{\tau (1 + \lambda_1 F_1)} + C_1,
\label{sfsol}
\end{equation}
where $C$ and $C_1$ are the integration constants.

Solving the Einstein equation for the metric functions we find
\begin{equation}
a_i(t) = A_{i} [\tau(t)]^{1/3} \exp \bigl[X_i \int [\tau
(t')]^{-1}dt' \bigr], \label{a_i}
\end{equation}
with the integration constants $A_i$ and $X_i$ obeying
$A_1 A_2 A_3 = 1$ and $X_1 + X_2 + X_3 = 0$. Note that to evaluate
the metric functions at any given time $\tilde{t}$ we should
first integrate $\int \frac{dt}{\tau}$, and only then
substitute $t$ by $\tilde{t}$.

The theoretical arguments~\cite{misner} and recent experimental
data which support the existence of an anisotropic phase that
approaches an isotropic one, led us to consider the models of
Universe with anisotropic background. On the other hand the isotropy
of the present-day Universe lead us to study how the initially
anisotropic BI space-time can evolve into an isotropic
Friedman-Robertson-Walker (FRW) one. Since for the FRW Universe
$a_1(t) = a_2(t) = a_3(t)$, for the BI universe to evolve into a
FRW one we should set $D_1 = D_2 = D_3 = 1$. Moreover, the
isotropic nature of the present Universe leads to the fact that
the three other constants $X_i$ should be close to zero as well,
i.e., $|X_i| << 1$, ($i = 1,2,3$), so that $X_i \int [\tau
(t)]^{-1}dt \to 0$ for $t < \infty$ (for $\tau (t) = t^n$ with $n
> 1$ the integral tends to zero as $t \to \infty$ for any $X_i$).
The rapid growth of the Universe due to the introduction of the
nonlinear spinor field to the system results in the earlier
isotropization.

As is seen from \eqref{psiI}, \eqref{psiJ}, \eqref{sfsol} and
\eqref{a_i}, the spinor, scalar and metric functions are in some
functional dependence of $\tau$. It should be noted that besides
these, other physical quantities such as spin-current, charge etc.
and invariant of space-time are too expressed via $\tau$
\cite{sprd,bited}. It should be noted that at any space-time
points where $\tau = 0$ the spinor, scalar and gravitational
fields become infinity, hence the space-time becomes singular at
this point \cite{bited}. So it is very important to study the
equation for $\tau$ (which can be viewed as master equation) in
details, exactly what we shall do in the section to follow. In
doing so we analyze the role of spinor field in the character of
evolution.

\section{Evolution of BI universe and role of spinor field}

In this section we study the role of spinor field in the evolution
of the Universe. But first of all let me qualitatively show the
differences that occur at the later stage of expansion depending
on how the sources of the gravitational field were introduced in
the system. In doing so we write the Einstein equation in the
following form:
\begin{subequations}
\label{ee}
\begin{eqnarray}
\frac{\ddot a_2}{a_2} +\frac{\ddot a_3}{a_3} + \frac{\dot a_2}{a_2
}\frac{\dot a_3}{a_3}&=&  \kappa T_{1}^{1} + \Lambda,\label{11}\\
\frac{\ddot a_3}{a_3} +\frac{\ddot a_1}{a_1} + \frac{\dot
a_3}{a_3}\frac{\dot a_1}{a_1}&=&  \kappa T_{2}^{2} +\Lambda,\label{22}\\
\frac{\ddot a_1}{a_1} +\frac{\ddot a_2}{a_2} + \frac{\dot a_1}{a_1
}\frac{\dot a_2}{a_2}&=&  \kappa T_{3}^{3} + \Lambda,\label{33}\\
\frac{\dot a_1}{a_1}\frac{\dot a_2}{a_2} +\frac{\dot
a_2}{a_2}\frac{\dot a_3}{a_3} +\frac{\dot a_3}{a_3}\frac{\dot
a_1}{a_1}&=& \kappa T_{0}^{0} + \Lambda. \label{00}
\end{eqnarray}
\end{subequations}
Here $\Lambda$ is the cosmological constant, $T_\mu^\nu$ is the
energy-momentum tensor of the source field. The Eq. \eqref{ee} is
thoroughly studied in \cite{sprd}. After a little manipulation
from \eqref{ee} one finds the equation for $\tau$ which is indeed
the acceleration equation and has the following general form:
\begin{equation}
\frac{\ddot \tau}{\tau}= \frac{3}{2}\kappa
\Bigl(T_{1}^{1}+T_{0}^{0}\Bigr) + 3 \Lambda, \label{dtau}
\end{equation}
Note also that here a positive $\Lambda$ corresponds to the
universal repulsive force which is often considered as a form of
dark energy, while a negative one gives an additional
gravitational force.

The Bianchi identity $G_{\mu;\nu}^{\nu}= 0$ in our case gives
\begin{equation}
{\dot T}_{0}^{0} = - \frac{\dot \tau}{\tau}\bigl(T_{0}^{0} -
T_{1}^{1}\bigr). \label{conservds}
\end{equation}
After a little manipulation from \eqref{dtau} and
\eqref{conservds} one finds the following expression for $T_0^0$:

\begin{equation}
\kappa T_0^0 = 3 H^2 - \Lambda - C_{00}/\tau^2, \label{kt00}
\end{equation}
where the definition of the generalized Hubble constant $H$
as
\begin{equation}
3H = \frac{\dot \tau}{\tau} = \frac{\dot a_1}{a_1}+ \frac{\dot
a_2}{a_2}+\frac{\dot a_3}{a_3} = H_1 + H_2 + H_3. \label{HC}
\end{equation}
Let us analyze the relation \eqref{kt00} in details. Consider the
case when $\Lambda = 0$. At the moment when the expansion rate is
zero (it might be at a time prior to the "Big Bang", or sometimes
in the far future when the universe cease to expand we have $H =
0$.) the nonnegativity of $T_0^0$ suggests that $C_{00} \le 0$.
Before considering the case for large $\tau$ we should like to
study the Eq. \eqref{conservds} in detail. For the spinor and
scalar fields chosen in this paper they are identically fulfilled.
If this is not the case, an additional equation, know as equation
of state, is applied to connect pressure ($T_1^1$) with energy density
($T_0^0$). In the long run from \eqref{conservds} one finds something
like $(T_0^0)^b \tau =$ const., where $b $ is some constant (in case of
perfect fluid $b = 1 + \zeta$). Thus we see that the energy density of
the source field introduced into the system as above decreases with the
growth of $\tau$. Now if we consider the case when $\tau$ is big
enough for $T_0^0$ to be neglected, from \eqref{kt00} we find
\begin{equation}
3 H^2 - \Lambda \to 0. \label{HL}
\end{equation}
On account of \eqref{HC} from \eqref{HL} one finds
\begin{equation}
\tau \to \exp{[\sqrt{3 \Lambda}\,t]}.  \label{tauHL}
\end{equation}
From \eqref{HL} and \eqref{tauHL} it follows that for $\tau$ to be
infinitely large, $\Lambda \ge 0.$ In case of $\Lambda = 0$ we
find that beginning from some value of $\tau$ the rate of
expansion of the Universe becomes trivial, that is the universe
does not expand with time. Whereas, for $\Lambda > 0$ the
expansion process continues forever.  As far as negative $\Lambda$
is concerned, its presence imposes some restriction on $\tau$,
namely, there exists some upper limit for $\tau$ (note that $\tau$
is essentially nonnegative, i.e. bound from below). Thus we see
that a negative $\Lambda$, depending on the choice of parameters
can give rise to an oscillatory mode of expansion \cite{sprd}.
Thus we can conclude the following :

{\it Let $T_\mu^\nu$ be the source of the Einstein field equation;
$T_0^0$ is the energy density and $T_1^1,\,T_2^2,\,T_3^3$ are the
principal pressure and $T_1^1 = T_2^2 = T_3^3$. An ever-expanding
BI Universe may be obtained if and only if the  $\Lambda$ term is
positive (describes a repulsive force and can be viewed as a form
of dark energy) and is introduced into the system as in \eqref{ee}
or if the source field introduced as a part of energy-momentum
tensor behaves like a $\Lambda$ term as $\tau \to \infty$.}

It should be noted that the sources of the gravitational field
such as spinor, scalar and electromagnetic fields, perfect or
imperfect fluids, as well as dark energy such as quintessence,
Chaplygin gas are introduced into the system as parts of the total
energy-momentum tensor $T_\mu^\nu$. It is also known that the dark 
energy was introduced into the system to explain the late time 
acceleration of the Universe. To show that though the dark energy 
is introduced into the system as a part of total energy-momentum 
tensor, it still behaves like a $\Lambda$ term as $\tau \to \infty$,
we write them explicitly. The quintessence and
Chaplygin gas are given by the following equation of states:
\begin{subequations}
\label{qc}
\begin{eqnarray}
p_{q} &=& w \ve_{q}, \quad w \in  [-1,0], \label{quint}\\
p_{c} &=& -A/\ve_{c}, \quad A > 0. \label{chap}
\end{eqnarray}
\end{subequations}
Note that the energy densities of the quintessence and Chaplygin
gas are related to $\tau$ as \cite{pfden}
\begin{subequations}
\label{qcs}
\begin{eqnarray}
\ve_q &=& \ve_{0q}/\tau^{1+w}, \quad w \in  [-1,0], \label{quints}\\
\ve_c &=& \sqrt{\ve_{0c}/\tau^2 + A}, \quad A > 0. \label{chaps}
\end{eqnarray}
\end{subequations}
From \eqref{qcs} and \eqref{qc} follows that $\ve_c \to
\sqrt{A}$ and $p_c \to - \sqrt{A}$ as $\tau \to \infty$. In case
of a quintessence, for $w > -1$, both energy density and pressure
tend to zero as $\tau$ tends to infinity. But for $w = -1$
(sometimes known as phantom matter) we have $\ve_q \to \ve_{0q}$
and $p_q \to - \ve_{0q}$ as $\tau \to \infty$. It means a
quintessence with $w = -1$ and Chaplygin gas behave like a
$\lambda$ term when $\tau \to \infty$ and hence can give rise
to an ever expanding Universe.

Before solving the equation for $\tau$ we have to write the
components of the energy-momentum tensor of the source fields 
in details:
\begin{eqnarray}
 T_{0}^{0} &=& mS - F + \frac{1}{2} ( 1 +  \lambda_1
F_1) {\dot \vf}^2 + \ve_{pf}, \nonumber\\ \label{total}
\\
T_{1}^{1} &=& T_{2}^{2} = T_{3}^{3} = {\cD} S + {\cG} P - F -
\frac{1}{2} ( 1 +  \lambda_1 F_1) {\dot \vf}^2  - p_{pf},\nonumber
\end{eqnarray}
where, ${\cD} = 2 S dF/dI + \lambda_1 S {\dot \vf}^2 dF_1/dI$ and
${\cG} = 2 P dF/dJ + \lambda_1 P {\dot \vf}^2 dF_1/dJ.$ In
\eqref{total} $\ve_{pf}$ and $p_{pf}$ are the energy density and
pressure of the perfect fluid, respectively and related by the
equation of state
\begin{equation}
p_{pf} = \zeta \ve_{pf}, \quad \zeta \in [0,\,1]. \label{eos}
\end{equation}

Let us now study the equation for $\tau$ in details and clarify
the role of material field in the evolution of the Universe. For
simplicity we consider the case when both $F$ and $F_1$ are the
functions of $I\, (S)$ only. We also set $C = 1$ and $C_0 = 1$.
Thanks to the spinor field equations and those for the invariants of
the bilinear spinor form, the energy-momentum conservation law 
for the spinor field satisfied identically \cite{sprd}. As a result 
the Eq. \eqref{conservds} now reads \cite{sprd}
\begin{equation}
\dot \ve + \frac{\dot \tau}{\tau} (\ve + p) = 0. \label{vepr}
\end{equation}
In view of \eqref{eos} from \eqref{vepr} for the energy density
and pressure of the perfect fluid one finds
$$\ve_{pf} = \frac{\ve_0}{\tau^{1+\zeta}}, \quad p_{pf} = \frac{\zeta_0
\ve_0}{\tau^{1+\zeta}}.$$ Further we set $\ve_0 = 1$. Assume that
$F = \lambda S^q$ and $F_1 = S^r$ where $\lambda$ is the
self-coupling constant. As it was shown in \cite{sprd}, the spinor
field equation, more precisely the equations for bilinear spinor
forms, in this case gives $S = C_0 /\tau$. Then setting $C_0 = 1$
for the energy density and the pressure from \eqref{total} we find
\begin{eqnarray}
 T_{0}^{0} &=& \frac{m}{\tau} - \frac{\lambda}{\tau^q} +
 \frac{\tau^{r-2}}{2(\lambda_1 + \tau^r)} +
 \frac{1}{\tau^{1 + \zeta}} \equiv \ve \, \nonumber\\ \label{total0}
\\
T_{1}^{1} &=& \frac{(q -1)\lambda}{\tau^q} - \frac{[(1 - r)
\lambda_1 + \tau^r ] \tau^{r-2}}{2(\lambda_1 + \tau^r)^2} -
\frac{\zeta}{\tau^{1+\zeta}} \equiv p.\nonumber
\end{eqnarray}

Taking into account that $T_0^0$ and $T_1^1$ are the functions of
$\tau$ only, the Eq. \eqref{dtau} can now be presented as
\begin{equation}
\ddot \tau = {\cal F}(q_1,\tau), \label{newtd}
\end{equation}
where we define
\begin{equation}
{\cal F}(q_1,\tau) =  (3/2) \kappa \Bigl(m + \lambda (q -2)\tau^{1
- q} + \lambda_1 r \tau^{r-1}/2(\lambda_1 + \tau^r)^2 +
(1-\zeta)/\tau^\zeta \Bigr), \label{force}
\end{equation}
where $q_1 = \{\kappa,m,\lambda,\lambda_1,q,r,\zeta\}$ is the set
of problem parameters. The En. \eqref{newtd} allows the
following first integral:
\begin{equation}
\dot \tau = \sqrt{2[E - {\cal U}(q_1,\tau)]} \label{1stint}
\end{equation}
where we denote
\begin{equation}
 {\cal U}(q_1,\tau) = - \frac{3}{2}\Bigl[\kappa\Bigl(m \tau -
 \lambda /\tau^{q - 2} - \lambda_1 /2(\lambda_1 + \tau^r) +
 \tau^{1-\zeta}\Bigr)\Bigr]. \label{poten}
\end{equation}
From a mechanical point of view Eq. \eqref{newtd} can be
interpreted as an equation of motion of a single particle with
unit mass under the force $\mathcal F(q_1,\tau)$. In
\eqref{1stint} $E$ is the integration constant which can be
treated as energy level, and ${\cal U}(q_1,\tau)$ is the potential
of the force $\mathcal F(q_1, \tau)$. We solve the Eq.
\eqref{newtd} numerically using Runge-Kutta method. The initial
value of $\tau$ is taken to be a reasonably small one, while the
corresponding first derivative $\dot \tau$ is evaluated from
\eqref{1stint} for a given $E$. 

Let us go back to the Eq. \eqref{newtd}. In view of \eqref{force} one sees, 
$\ddot \tau \to (3/2) \kappa m > 0$ as $\tau \to \infty$, i.e., if 
$\ddot \tau$ is considered to be the acceleration of the BI Universe, 
then the massive spinor field essentially can be viewed as a source for ever 
lasting acceleration. Note that it does not contradicts our previous statement 
about the role of energy-momentum tensor on ever expanding Universe, since
the spinor field satisfies the Bianchi identity identically.

Now a few words about considering $\ddot \tau$ as acceleration. 
The Einstein equations for the FRW model read
\begin{subequations}
\label{frw}
\begin{eqnarray}
2 \frac{\ddot a}{a} + \Bigl(\frac{\dot a}{a}\Bigr)^2 &=& \kappa
T_1^1, \label{FRW11} \\
3 \Bigl(\frac{\dot a}{a}\Bigr)^2 &=& \kappa T_0^0. \label{FRW00}
\end{eqnarray}
\end{subequations}
From \eqref{frw} one finds
\begin{equation}
\frac{\ddot a}{a} = - \frac{\kappa}{6} (T_0^0 - 3 T_1^1), \label{accfrw} 
\end{equation}
The equation \eqref{accfrw} is known as the acceleration equation. In analogy
for the BI Universe from \eqref{ee} we can write
\begin{equation}
\frac{\ddot a_1}{a_1} + \frac{\ddot a_2}{a_2} + \frac{\ddot a_3}{a_3} = 
- \frac{\kappa}{2} (T_0^0 - 3 T_1^1), \label{accbi1} 
\end{equation}
and declare it as acceleration equation. Though setting $a_1 = a_2 = a_3$ 
we recover the original definition, hardly it will be helpful
in our case. So in BI Universe we assume $\ddot \tau$ be the acceleration and
Eq. \eqref{dtau} be the acceleration equation.   

Let us now define the deceleration parameter. In FRW cosmology
the deceleration parameter has the form
\begin{equation}
d_{\rm frw} = - \frac{a \ddot{a}}{{\dot a}^2} = - 
\Bigl[1 + \frac{\dot H_{\rm frw}}{H_{\rm frw}}\Bigr] =
\frac{d}{dt}\bigl(\frac{1}{H_{\rm frw}}\bigr) -1, \label{decprfrw}
\end{equation}
where $H_{\rm frw} = \dot{a}/a$ is the Hubble parameter for FRW model.
In analogy we can define a deceleration parameter as well. If we 
define the generalized deceleration parameter in the following way:
\begin{equation}
d = -\Bigl[1 + \frac{{\dot H_1} + {\dot H_2} + {\dot H_3}}{H_1^2 +
H_2^2 + H_3^2}\Bigr], \label{decpo}
\end{equation}
where $H_i = \dot{a}_i/a_i$, then the standard deceleration parameter 
is recovered at $a_1 = a_2 = a_3$.  But is this case the definition for
acceleration adopted here is no longer valid. So we switch to the second 
choice and following Belinchon and Harko {\it et. al.} \cite{belin,harko}
define the generalized deceleration parameter as
\begin{equation}
d = \frac{d}{dt} \bigl(\frac{1}{3H}) - 1 = - \frac{\tau \ddot \tau}{{\dot \tau}^2}.
\label{decp}
\end{equation}
After a little manipulation in view of \eqref{ee} and \eqref{kt00} the
deceleration parameter can be presented as
\begin{equation}
d = -\frac{\kappa}{2} \frac{(T_1^1 + T_0^0) \tau^2}{\kappa T_0^0
\tau^2 +  C_{00}} \label{decr}
\end{equation}

Let us now go back to the equations \eqref{newtd}, \eqref{force}, \eqref{1stint}
and \eqref{poten}. As one sees, the positivity of the
radical imposes some restriction on the value of $\tau$, namely in
case of $\lambda > 0$ and $q \ge 2$ the value of $\tau$ cannot be
too close to zero at any space-time point. It is clearly seen from
the graphical view of the potential [cf. Fig. \ref{potp}]. Thus
we can conclude that for some special choice of problem parameters
the introduction of nonlinear spinor field given by a self-action
provides singularity-free solutions. As it was shown in
\cite{sprd} the regular solution is obtained only at the expense
of broken dominant-energy condition in the Hawking-Penrose
theorem.

{\it If, in an eigentetrad  of $T_{\mu \nu}$, $\ve$ denotes the
energy density and $p_1,\,p_2,\,p_3$ denote the three principal
pressure, then the dominant energy condition can be written as}
\cite{hawking}:
\begin{subequations}
\begin{eqnarray}\label{dec}
\ve + \sum_{\alpha} p_\alpha &\ge& 0;\\
\ve + p_\alpha &\ge& 0,
\quad \alpha = 1,2,3.
\end{eqnarray}
\end{subequations}
The dominant energy condition for the BI metric can be written in
the form:
\begin{subequations}
\begin{eqnarray}
T_{0}^{0} &\ge& T_{1}^{1} a_1^2 + T_{2}^{2} a_2^2 + T_{3}^{3} a_3^2,\\
T_{0}^{0} &\ge& T_{1}^{1} a_1^2, \\ T_{0}^{0} &\ge& T_{2}^{2} a_2^2,\\
T_{0}^{0} &\ge& T_{3}^{3} a_3^2.
\end{eqnarray}
\end{subequations}

In Fig. \ref{potm} we plot the potential for a negative $\lambda$. As one sees,
in the vicinity of $\tau = 0$ there exists a bottomless potential hole. As one sees, 
if in case of a self-action the initial value of $\tau$ is too close to zero and 
the constant $E$ is less than ${\cal U}_{\rm max}$ (the maximum value of the potential
in presence of a self-action), the Universe will never come out of the hole.

For numerical solutions we set $\kappa = 1$, spinor mass $m = 1$,
the power of nonlinearity we choose as  $q = 4$, $ r = 4$ and for
perfect fluid we set $\zeta = 1/3$ that corresponds to a
radiation. We also set $C_{00} = -0.001$ and $E = 10$. The initial
value of $\tau$ is taken to be $\tau_0 = 0.4$. The coupling constant
is chosen to be $\lambda_1 = 0.5$, while the self coupling constant
is taken to be either $\lambda = 0.5$ or $\lambda = -0.5$.
Here, in the figures we use the following notations:\\
1 corresponds to the case with self-action and interaction;\\ 
2 corresponds to the case with self-action only;\\
3 corresponds to the case with interaction only.

\vskip 1 cm

\myfigures{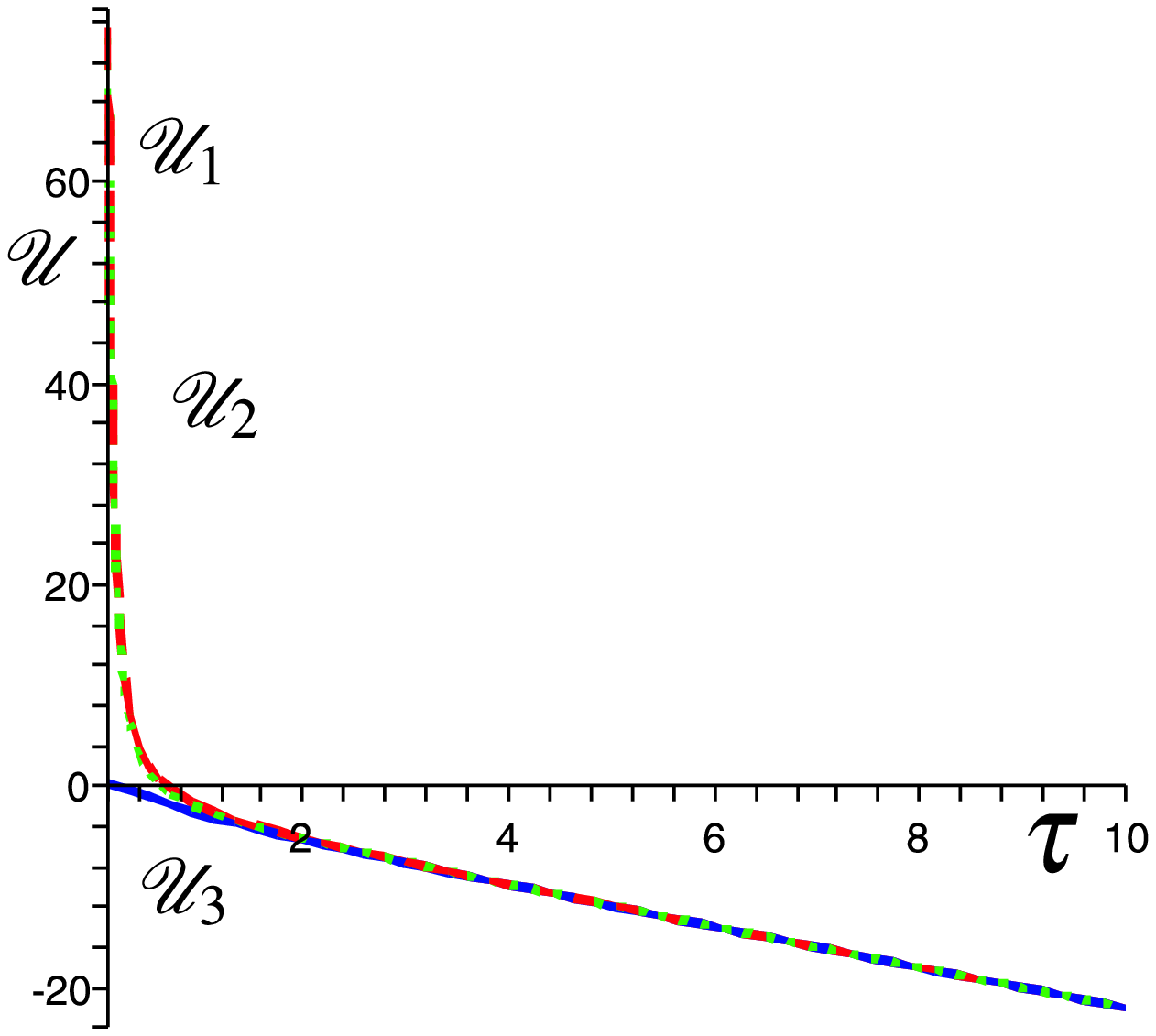}{0.45}{View of the potential $\mathcal U(\tau)$
for $\lambda > 0$.} {0.45}{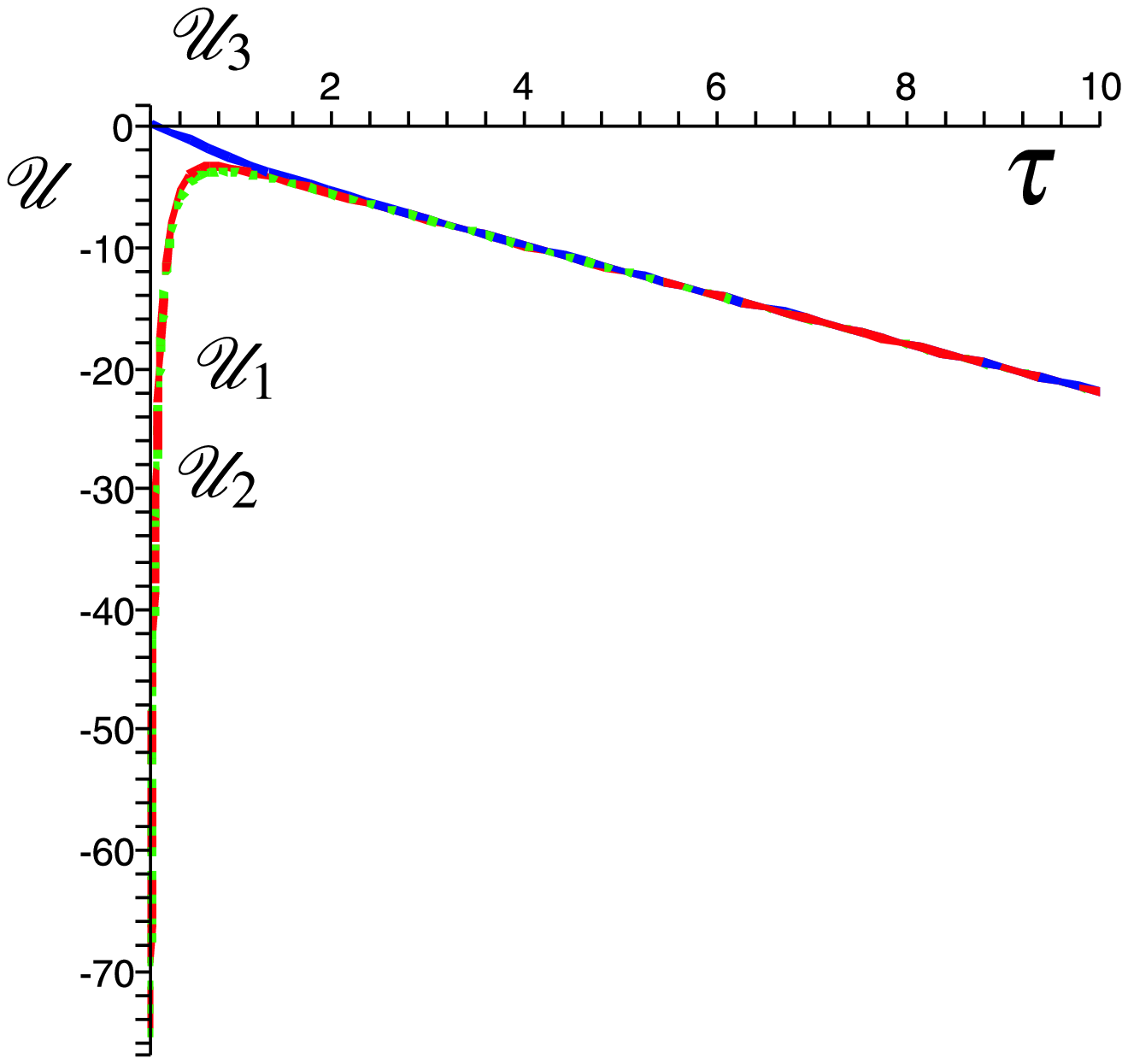}{0.45}{View of the potential 
$\mathcal U(\tau)$ for a negative $\lambda$.}{0.45}

As one sees from Fig. \ref{potp}, in presence of a self-action
of the spinor field with a positive $\lambda$, there occurs an 
infinitely high barrier as
$\tau \to 0$, it means that in the case considered here $\tau$
cannot be trivial [if treated classically, the Universe cannot
approach to a point unless it stays at an infinitely high energy
level]. Thus, the nonlinearity of the spinor field provided
by the self-action generates singularity-free evolution of the
Universe. But, as was already mentioned, this regularity can be
achieved only at the expense of dominant energy condition in
Hawking-Penrose theorem. It is also clear that if the nonlinearity
is induced by a scalar field, $\tau$ may be trivial as well, thus
giving rise to space-time singularity. Note that cases in presence
of a $\Lambda$ term are thoroughly studied in \cite{sprd,bited}.
It was shown that introduction of a positive $\Lambda$ just
accelerates the speed of expansion, whereas, a negative $\Lambda$
depending of the choice of $E$ generates oscillatory or
non-periodic mode of evolution. Note also that the regular
solution obtained my means of a negative $\Lambda$ in case of
interaction does not result in broken dominant energy condition
\cite{bited}.

In Figs. \ref{epp} and \ref{epm} we plot the corresponding energy density
and pressure. As one sees, in case of a positive $\lambda$ the energy density
is initially negative while the pressure is positive. In this case though
the solution is singularity-free, the violation of dominant energy takes place. 
In case of a negative $\lambda$ the pressure is always negative.

\myfigures{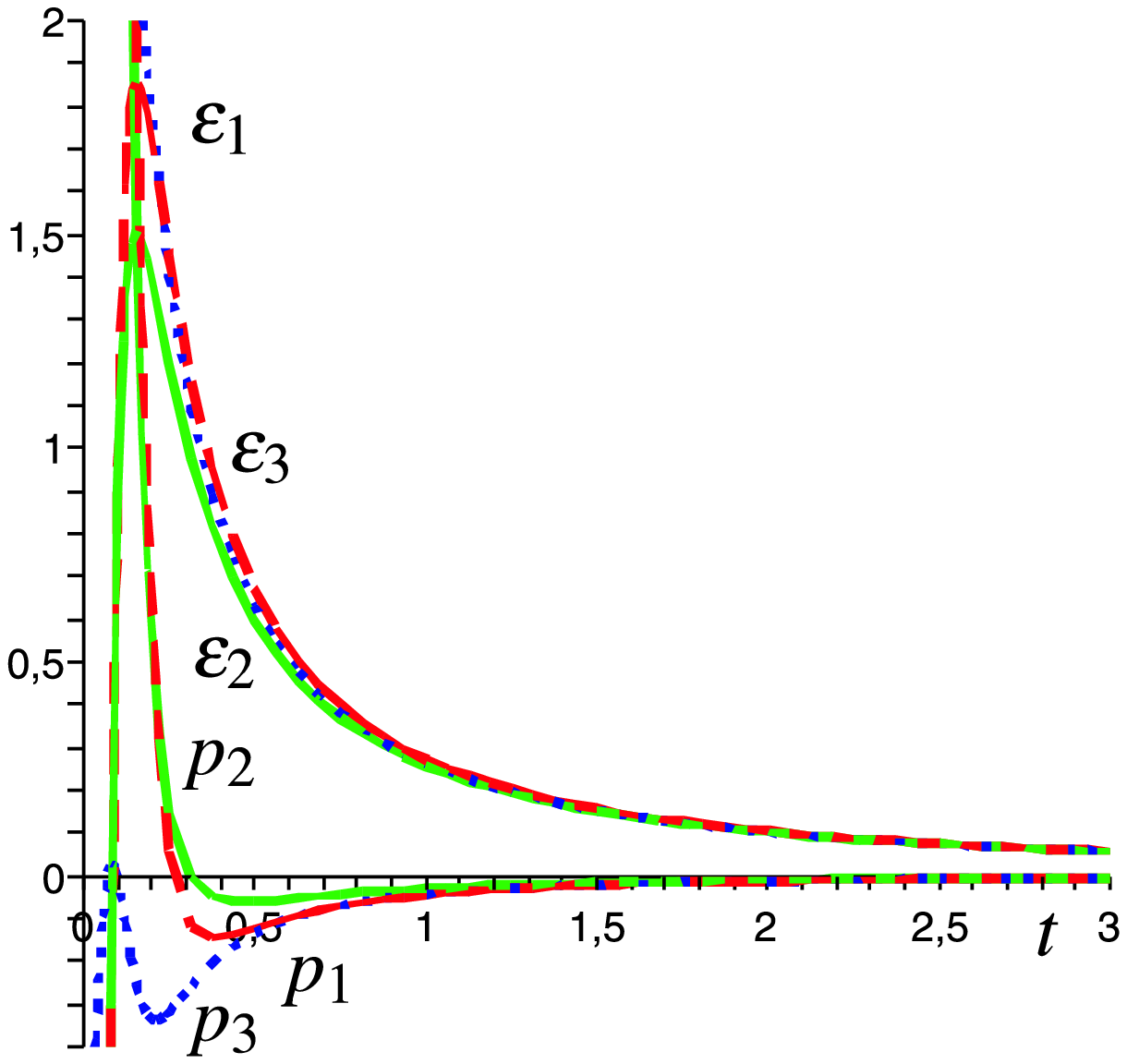}{0.45}{Energy density and pressure corresponding
to a positive $\lambda$.} {0.45}{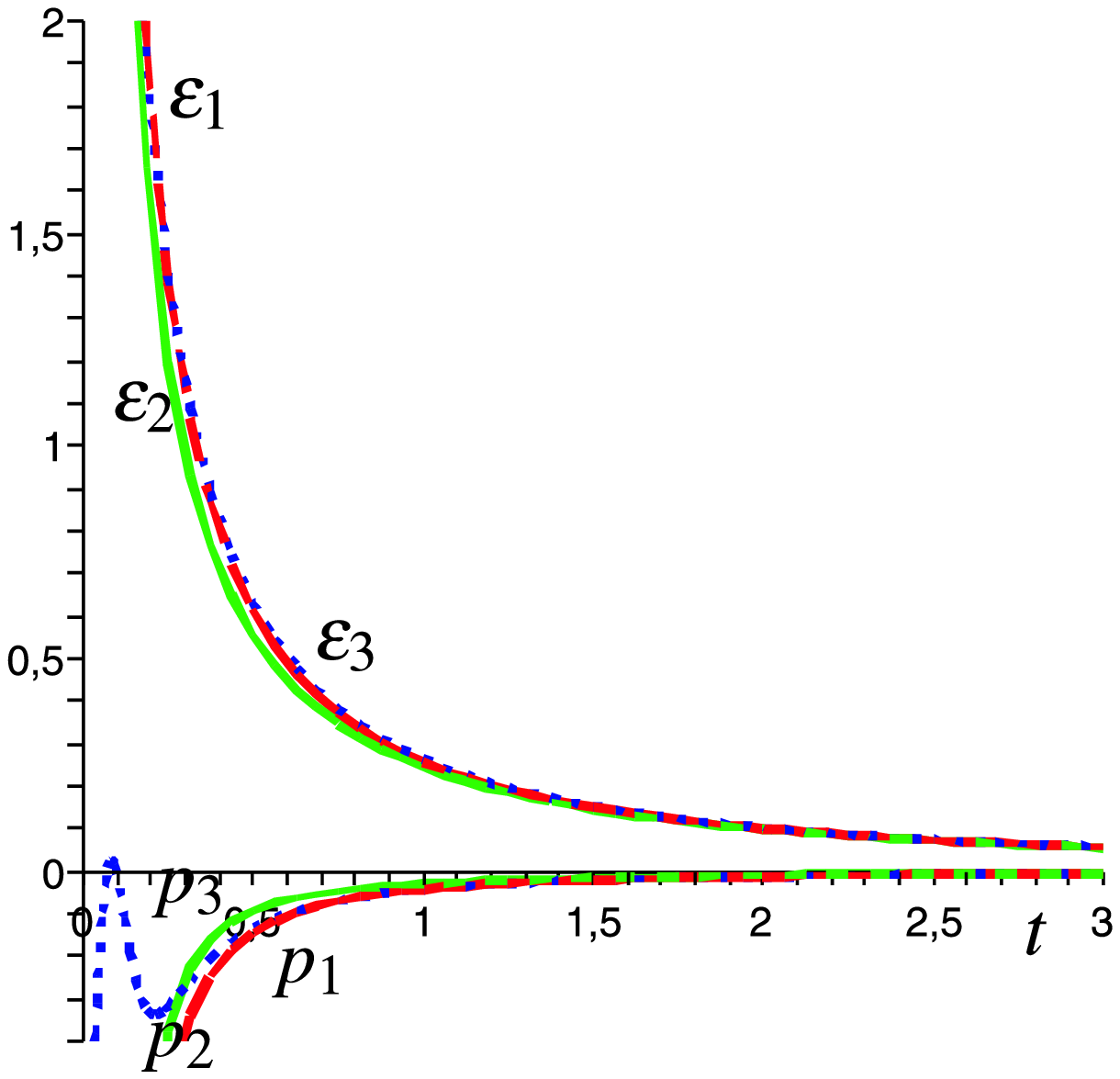}{0.45}{Energy density and
pressure in case of a negative $\lambda$.}{0.45}

The purpose of plotting the energy density and pressure is to show 
that the energy density of the source field indeed decreases with 
the increase of the Universe. This also shows that there exists an 
interval where the
energy density of the system with spinor field nonlinearity
generated by the self-action is negative. This is in line with our
earlier assumption. Moreover, we see the pressure of the source
field becomes negative in course of evolution (In case of
self-action with a positive $\lambda$ pressure is initially positive, 
but with the expansion of the Universe it becomes negative, whereas, 
in case of of a negative $\lambda$ as well as in case of 
interacting fields the pressure is always negative). Recall that
the dark energy (e.g. quintessence, Chaplygin gas), modelled to
explain the late time acceleration of the Universe, has the
negative pressure. So we argue that the models with nonlinear spinor field 
and interacting spinor and scalar fields to some extent can be considered as an 
alternative to dark energy which is able to explain the late time 
acceleration of the Universe.

In the Figs. \ref{acp} and \ref{acm} we illustrate the acceleration
of the Universe for positive and negative $\lambda$, respectively.
As one sees, in both cases we have decreasing acceleration that 
tends to $(3/2) \kappa m$ as $\tau \to \infty$.

\myfigures{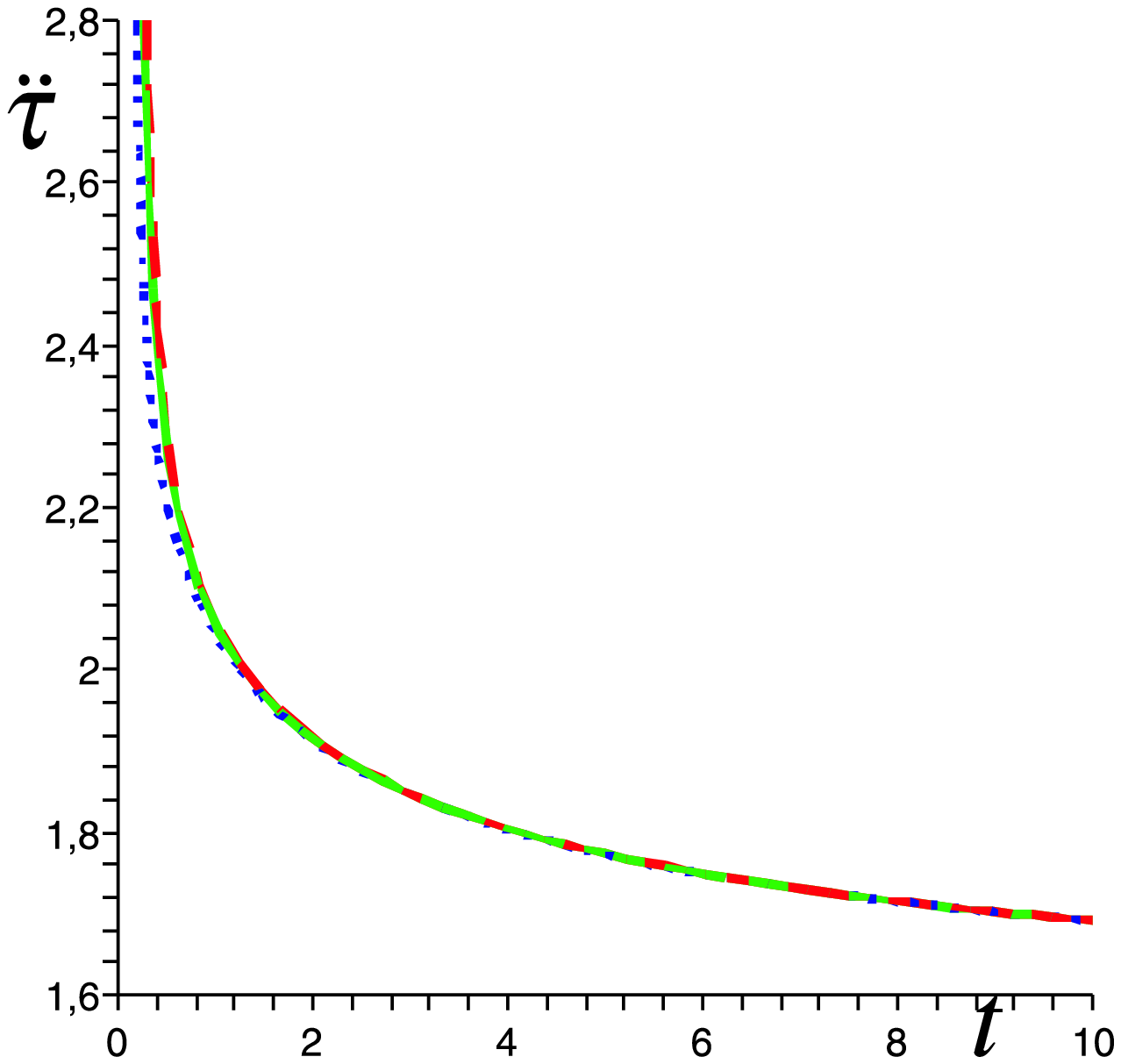}{0.45}{Acceleration of the Universe corresponding
to a positive $\lambda$.} {0.45}{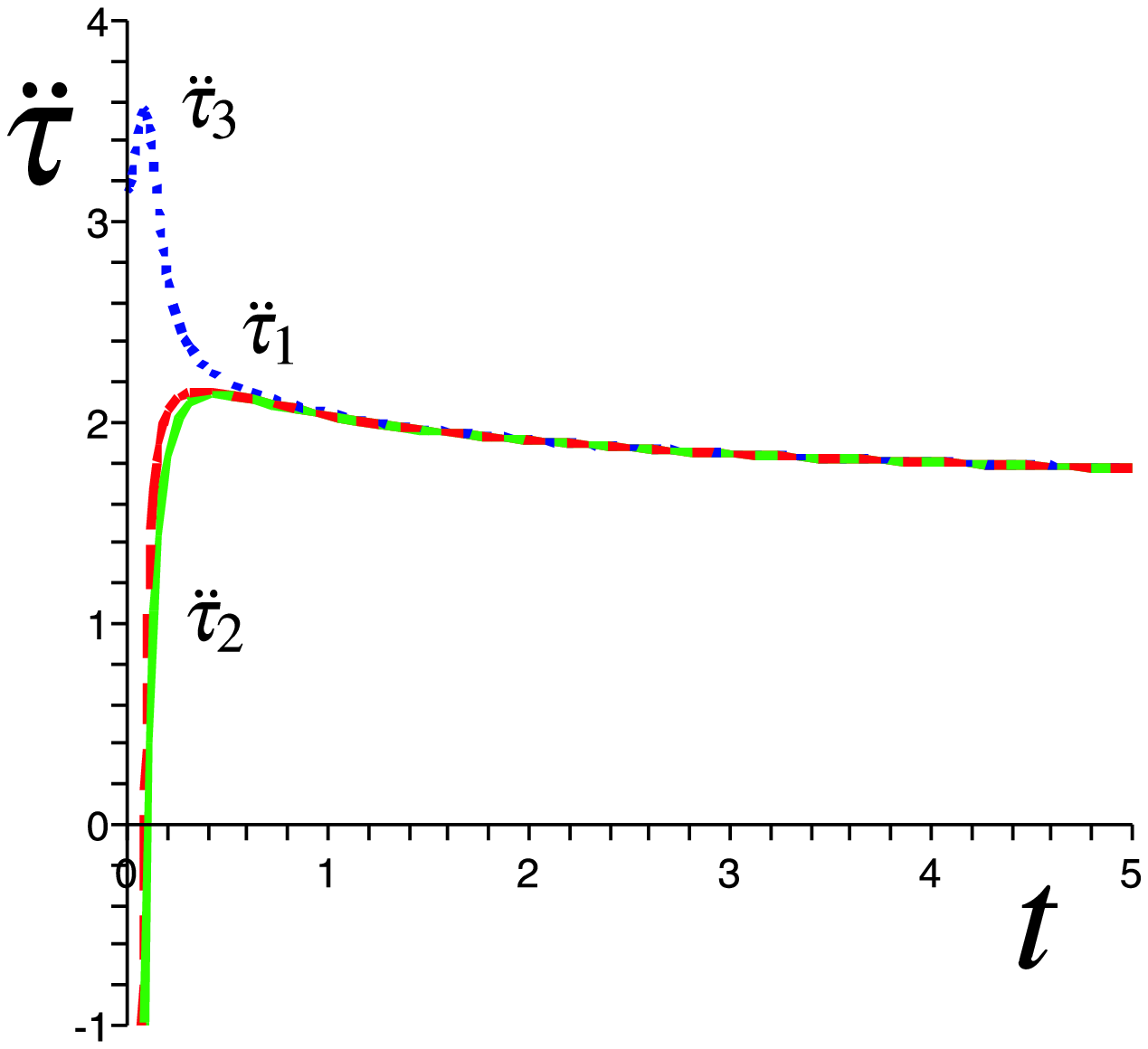}{0.45}{Acceleration of the 
Universe in case of a negative $\lambda$.}{0.45}

In Figs. \ref{dpp} and \ref{dpm} we plot the deceleration parameter.

\myfigures{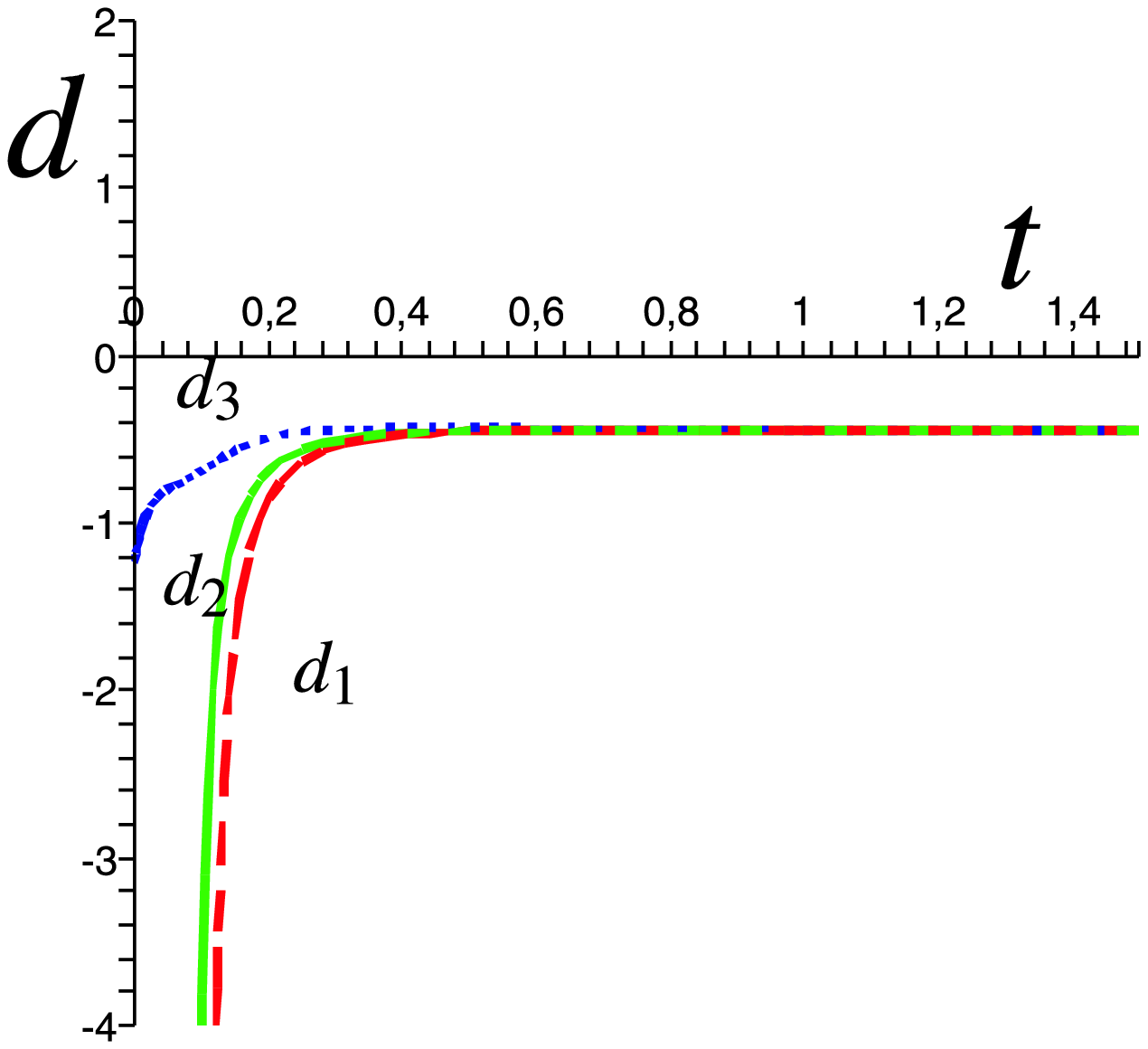}{0.45}{Deceleration parameter corresponding
to a positive $\lambda$.} {0.45}{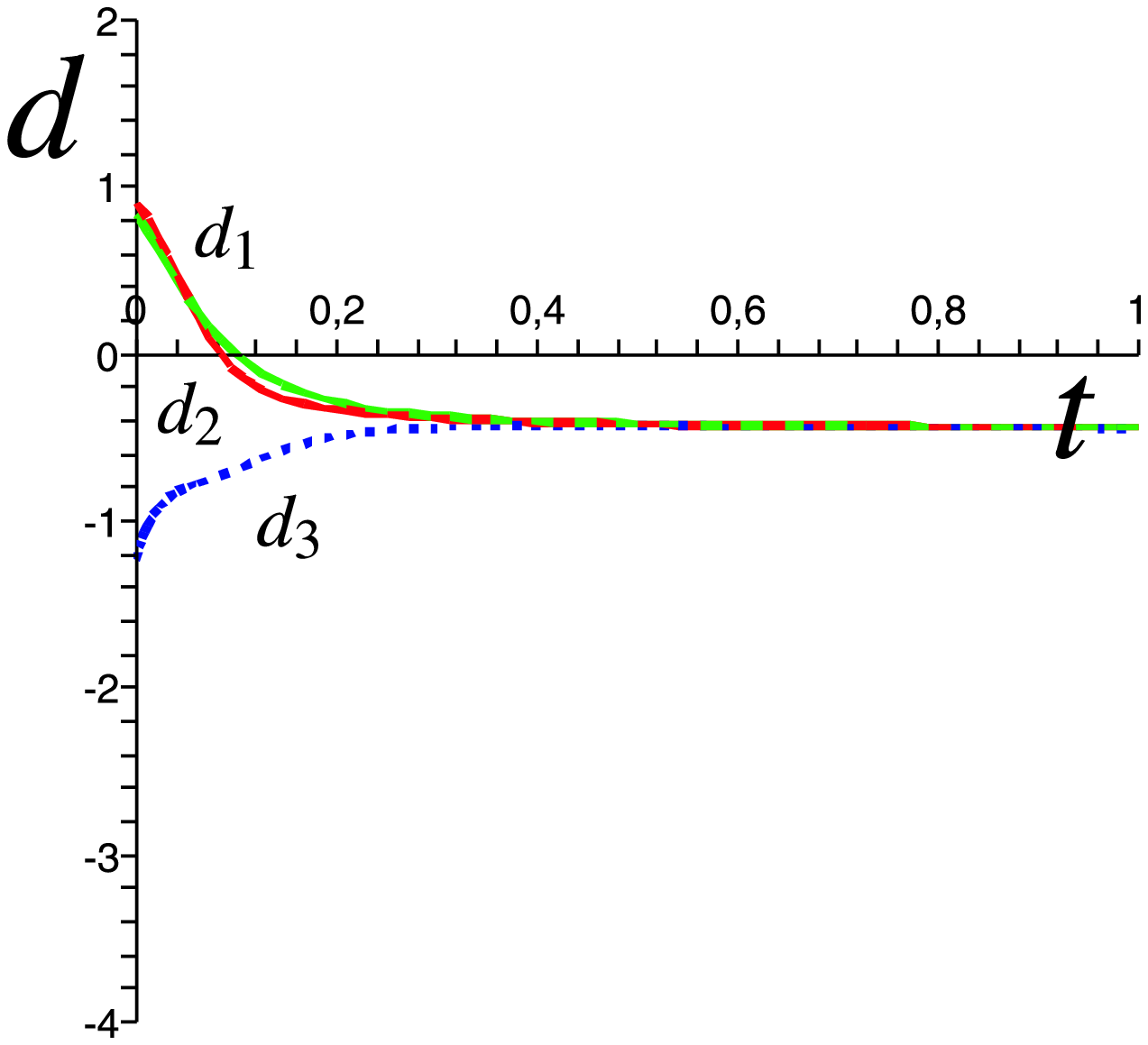}{0.45}{Deceleration parameter 
in case of a negative $\lambda$.}{0.45}

The Figs. \ref{acp}, \ref{acm}, \ref{dpp} and \ref{dpm} show the
accelerated mode of expansion of the Universe. As one sees, the
acceleration is decreasing with time. Depending of the choice of
nonlinearity it undergoes an initial deceleration phase. It is also
seen that the nonlinear term plays proactive role at the initial stage 
while at the later stage spinor mass is crucial for the accelerated
mode of expansion.

\section{conclusion}

We considered a system of interaction nonlinear spinor and scalar
fields within the scope of a BI cosmological model filled with
perfect fluid. The spinor field nonlinearity gives rise to an effective 
negative pressure in the course of evolution. Comparison of the effective 
pressure of the nonlinear spinor field with that of a dark energy given
by a quintessence or Chaplygin gas leads us to conclude that the spinor 
field can be seen as an alternative to the dark energy able to explain the
acceleration of the Universe. It was shown that the nonlinear spinor term
is proactive at the early stage of the evolution and essentially accelerates
the process of evolution, while at the later stage of evolution the spinor 
mass holds the key. Given the fact that {\it  neutrino} is described by the 
spinor field equation and it too possesses mass (though too small but nonzero), 
the presence of huge number of neutrino in the Universe can be seen as one of 
the possible factor of the late time acceleration of the Universe. It was also
shown that for some specific choice of parameters it is possible to construct
singularity-free model of the Universe, but this regularity results in the broken
dominant energy condition of the Hawking-Penrose theorem.

\end{document}